# Organic Single Crystal Phototransistors: Recent Approaches and Achievements

Changbin Zhao[1], Muhammad Umair Ali[2], Jiaoyi Ning[1], Hong Meng[1,3]*

[1]School of Advanced Materials, Peking University Shenzhen Graduate School, Shenzhen 518055, China.

[2]Tsinghua-Berkeley Shenzhen Institute (TBSI), Tsinghua University, Shenzhen, 518055, China

[3]Frontiers Science Center for Flexible Electronics, Xi'an Institute of Flexible Electronics (IFE) and Xi'an Institute of Biomedical Materials & Engineering, Northwestern Polytechnical University, Xi'an, 710072, China.

*Corresponding author. Email: menghong@pku.edu.cn (H.M.).

**Abstract**

Organic phototransistors (OPTs), compared to traditional inorganic counterparts, have attracted a great deal of interest because of their inherent flexibility, light-weight, easy and low-cost fabrication, and are considered as potential candidates for next-generation wearable electronics. Currently, significant advances have been made in OPTs with the development of new organic semiconductors and optimization of device fabrication protocols. Among various types of OPTs, small molecule organic single crystal phototransistors (OSCPTs) standout because of their exciting features, such as long exciton diffusion length and high charge carrier mobility relative to organic thin-film phototransistors. In this review, a brief introduction to device architectures, working mechanisms and figure of merits for OPTs is presented. We then overview recent approaches employed and achievements made for the development of OSCPTs. Finally, we spotlight potential future directions to tackle the existing challenges in this field and accelerate the advancement of OSCPTs towards practical applications.

**Keywords:** photodetectors, organic transistors, organic phototransistors, organic single crystals, flexible electronics

# 1 Introduction

Photodetector are essential optoelectronic device that can detect light and convert it into electrical signals,[1-5] and offers a wide range of applications in various fields, such as image and biomedical sensing, video imaging, optical communication, night vision and health monitoring.[6-9] Depending on the working mechanism, photodetectors are generally categorized into three types: photoconductors, photodiodes and phototransistors.[10,11] Phototransistors, integrate light capture capability and signal amplification function in a single device, and exhibit higher optical detectivity and lower noise as compared to that of photoconductors and photodiodes.[12] The concept of phototransistors was first proposed by William Shockley in 1951.[13] To date, most of the commercial photodetectors employ phototransistors based on inorganic semiconductors due to their high charge-carrier mobility, small exciton binding energy and high stability. However, the practical applications of inorganic photodetectors are significantly limited due to a variety of drawbacks, including complicated and expensive manufacturing processes, and poor mechanical flexibility. In this scenario, organic phototransistors emerge as a potential competitor with inspiring features such as high flexibility, stretchability as well as simple and low-cost fabrication, making them suitable for application in next-generation wearable electronic devices.[14-19]

With the rapid development of organic materials and device fabrication technology, the mobility of organic semiconductors now exceeds that of polysilicon,[20,21] thereby making them able to meet the industrial requirements. Great achievements have also been made in phototransistors based on organic small-molecule semiconductors with R reaching up to 27000 A $W^{-1}$, which is comparable to that of inorganic counterparts.[22] Owing to the advantage of long exciton diffusion length and high charge carrier mobility, small-molecule organic single crystal phototransistors (OSCPTs) have attracted considerable attention, and their performance has been significantly improved. Furthermore, single crystal-based device is an ideal system for studying the molecular arrangement, charge carrier migration and photoelectric conversion due to long-range order and fewer defects in single crystals. In recent years, with the synthesis of new materials and progress of single crystal preparation technology, the photoelectric properties of organic single crystal transistors have been deeply investigated.[23,24]

In this review, we first introduce the basic structures, working mechanism and fundamental parameters of phototransistors. Then, recent advancement in OSCPTs classified by their architectures is spotlighted. Finally, we provide a summary and an outlook of OSCPTs for future development. This review will offer the readers a comprehensive understanding on OSCPTs, that would accelerate the development of this exciting technology.

## 2 Fundamentals of organic phototransistors (OPTs)

First, to clarify the basic knowledge of OPTs, in this section, we introduce typical device architectures, working principles and photoelectrical characteristics.

### 2.1 Device architectures

The configuration of phototransistors can be divided into five typical structures: bottom-gate/top-contact (BGTC), bottom-gate/bottom contact (BGBC), top-gate/top-contact (TGTC), top-gate/bottom-contact (TGBC), and vertical structure, as shown in Fig. 1, like the classification of organic field-effect transistors (OFETs).[4] The only difference of OPTs from OFETs is the external light source. In order to achieve maximum photoelectric performance, the light should be irradiated on the semiconductor channel directly and vertically, therefore, if the channel is enclosed (as in the case of TGTC and TGBC structures), the outer layer needs to be transparent enough so that the light could efficiently arrive at the channel to obtain more light-generated carriers and hence, high photo-detection performance. Since the active layer is directly exposed to the light source for phototransistors with BGTC and BGBC architectures, there is no special requirement compared with ordinary OFETs while they could be employed for a broader range of applications. Moreover, phototransistors with vertical structure can obtain high output current at low driving voltage due to their shorter channel length. In conclusion, each type of device structures differently affects the resulting performance of the phototransistors.

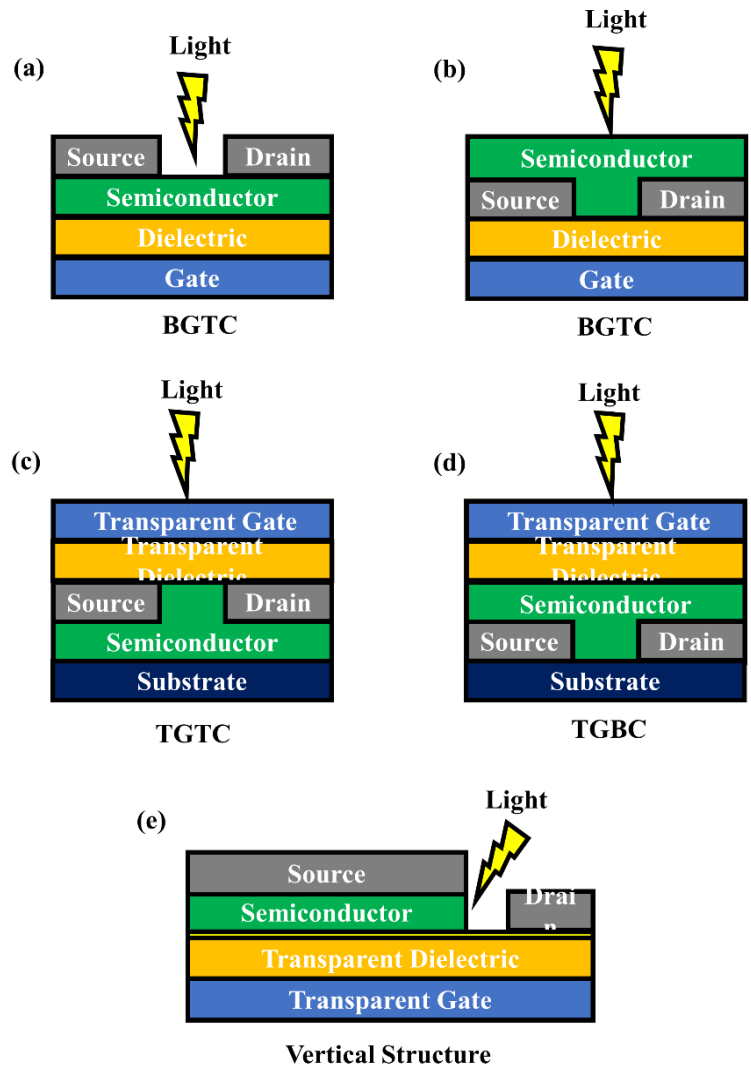

**Fig. 1** Typical device architectures for OPTs and light illumination: (a) BGTC, (b) BGBC, (c)TGTC, (d) TGBC and (e) vertical structure.

## 2.2 Working principles

In general, working principle of OPTs can be divided into two types according to the different exciton separation modes by on/off state of the transistors: (i) photovoltaic mode, and (ii) photoconductive mode. The details are as follows.

**Photovoltaic mode:** When an OPT is in on-state and works in the accumulation mode,

photovoltaic effect plays a major role in photo-detection. For a p-type semiconductor, the photoexcited holes transport towards the drain electrode while electrons accumulate in source electrode, which effectively reduces the contact resistance of hole injection in the source electrode. Hence, the threshold voltage ($V_{th}$) shifts towards more positive values and the drain current ($I_d$) under light illumination would be higher than that in dark under the same gate voltage ($V_g$). By contrast, the $V_{th}$ shifts towards more negative values and $I_d$ increases under light illumination due to the photovoltaic effect for an n-type semiconductor.[25-27] The photocurrent resulting from the photovoltaic effect can be modeled according to Equation (1):

$$I_{ph,pv} = g_m \Delta V_{th} = \frac{AkT}{q} \ln\left(1 + \frac{\eta q \lambda P_{opt}}{I_{dark} hc}\right) \quad (1)$$

where $g_m$ is the transconductance, $\Delta V_{th}$ is the threshold shift, A is a proportionality parameter, k corresponds to the Boltzmann constant, T signifies the temperature, $\eta$ is the photo-generation quantum efficiency, q refers to the elementary charge, $\lambda$ is the wavelength of light, $P_{opt}$ is the incident optical power in the channel area, $I_{dark}$ is the dark current for minority charges, and $hc/\lambda$ is the photon energy. Obviously, the photocurrent resulting from the photovoltaic effect increases exponentially with the incident optical power.

**Photoconductive mode:** When an OPT is in off-state and operates in the depletion mode, the photoconductive effect is dominant of $I_d$ increasing under light illumination. In fact, the conductivity and $I_d$ of organic semiconductors increase proportionally with an increase in the light intensity.[28,29] The photocurrent caused by photoconductive effect can be expressed by Equation (2):

$$I_{ph,pc} = (q\mu_p pE)WD = BP_{opt} \quad (2)$$

where $\mu_p$ is the majority charge carrier's mobility, p is the charge carrier concentration, E refers to the electric field in the conducting channel, W is the gate width, D corresponds to the depth of absorption region, and B is a proportionality factor.

### 2.3 Characterization

OPTs are fundamental optoelectronic devices that convert optical signals into electrical signal. Furthermore, the signal can be amplified by the multiplier effect of the transistor. To evaluate the photo-detection performance, external quantum efficiency (EQE), photoresponsivity (R), photosensitivity (P) and detectivity (D) are very important parameters; the details are provided below.

**External quantum efficiency (EQE):** EQE is the ratio of photo-generated carriers to the number of incident photon and describes the photoelectric conversion capability of an OPT. EQE of OPTs is generally greater than 100% because of the multiplier effect. It can be expressed as Equation (3):[14]

$$EQE = \frac{I_{light}/q}{P_{opt}/hv} \quad (3)$$

where $I_{light}$ is the photocurrent, q is the elementary charge, h is the Planck's constant, and v is the frequency of incident light.

**Responsivity (R):** R is defined as the ratio of photocurrent to incident-light intensity, which is usually given by the following relation Equation (4):[14]

$$R = \frac{I_{light} - I_{dark}}{P_{opt}} \quad (4)$$

R is used to indicate how efficiently a photodetector responds to a given incident optical signal.

**Photosensitivity (P):** P is the ratio of light to dark current, which is mathematically written as Equation (5):[11]

$$P = \frac{I_{light} - I_{dark}}{I_{dark}} \quad (5)$$

**Detectivity (D):** D is used to describe the ability of weak light detection and its normalized value is called specific detectivity (D*), which can be written as Equation (6):[11]

$$D^* = \frac{R\sqrt{A}}{\sqrt{2qI_{dark}}} \quad (6)$$

where A is the channel area, the higher D* means better detection of weak light.

**Response time (t):** Response time represents the speed of response to a rapidly modulated light signal. It is defined as the time from dark current to photocurrent (the rise time: $t_r$) or photocurrent to dark current (the falling time: $t_f$) (or rise from 10 to 90% or to decay from 90 to 10% of the peak value).[30] The shorter response time means more quick response to the incident light; it is a vital parameter when phototransistors are integrated into circuits and act as dynamic response photodetectors.

All of the above provided formulas for various parameters are interrelated and can be derived from each other. For example, R can be given by EQE and expressed as Equation (7):[11]

$$R = \frac{EQE \cdot \lambda q}{hc} \qquad (7)$$

## 3. Organic single crystal phototransistors (OSCPTs)

Single crystal transistors exhibit intrinsic properties of organic materials and have long exciton diffusion length with high charge carrier mobility relative to thin-film devices due to their ordered arrangement and fewer defects.[31,32] Therefore, fabricating OSCPTs is an ideal approach to explore the mechanism of OPTs and a promising choice to realize high-performance photodetectors.[3,33] Owing to the inspiring potential of OSCPTs, a lot of research efforts have been dedicated to develop these devices in recent years. In this section, a brief introduction to OSCPTs is provide followed by an overview of recent advancements in this area based on their device architectures.

### 3.1 Single component OSCPTs

Single component OSCPTs are comprised of a neat arrangement of molecules without the influence of grain boundaries. On one hand, these devices can be used to study the internal charge transport mechanism of organic semiconductors due to their intrinsic photoelectric properties. On the other hand, single crystal devices can obtain a higher charge mobility and photoresponsivity than the polycrystalline equivalents, thereby making them of great significance for the future industrial applications. Moreover, due to the single component and narrow absorption spectrum, OSCPTs often exhibit good response to a specific wavelength of light. Wu and co-authors[34] prepared C8-BTBT OSCPTs by solution process which exhibited ultrahigh response to visible-blind and deep-ultraviolet (UV) signals because of the admirable charge transport ability of the C8-BTBT single-crystals[Fig. 2(a)]. Particularly, in the case of very weak 365 nm UV radiation (0.2 mW cm$^{-2}$), P and R of their devices reached to $3.0 \times 10^4$ and 1200 A W$^{-1}$, whereas, at 1 mW cm$^{-2}$ 280 nm deep-UV light, the P and R values of 8300 and 44 A W$^{-1}$ were obtained, respectively. Zhao and colleagues[35] also synthesized a new molecule, BBDTE and fabricated BBDTE-based OSCPTs, which showed a carrier mobility of up to 1.62 cm$^2$V$^{-1}$s$^{-1}$ and exhibited a maximum R value approaching 9821 A W$^{-1}$ with a maximum P of about $10^5$ towards UV light (at weak light intensity of 37 μW cm$^{-2}$) while the P observed within tested light intensity range was unexpectedly stable[Fig. 2(b)]. Moreover, Tao et al.[36] reported UV-sensitive phototransistors based on two novel molecules, 1,6-DTEP and 2,7-DTEP single crystals grown by solution evaporation method and the corresponding devices manifested a maximum hole mobility of up to 2.1 cm$^2$V$^{-1}$s$^{-1}$. Also, OSCPTs based on 1,6-DTEP and 2,7-DTEP exhibited ultra-sensitivity to UV light with an ultrahigh R of up to $2.86 \times 0^6$ and $1.04 \times 10^5$ A W$^{-1}$ with corresponding D* of as high as $1.49 \times 10^{18}$ and $5.28 \times 10^{16}$ Jones, respectively, [Fig. 2(c)] which is ascribed to the presence of triphenylamine group that offers high photoluminescence quantum yields and enhances the charge carrier mobility of resulting single crystals.

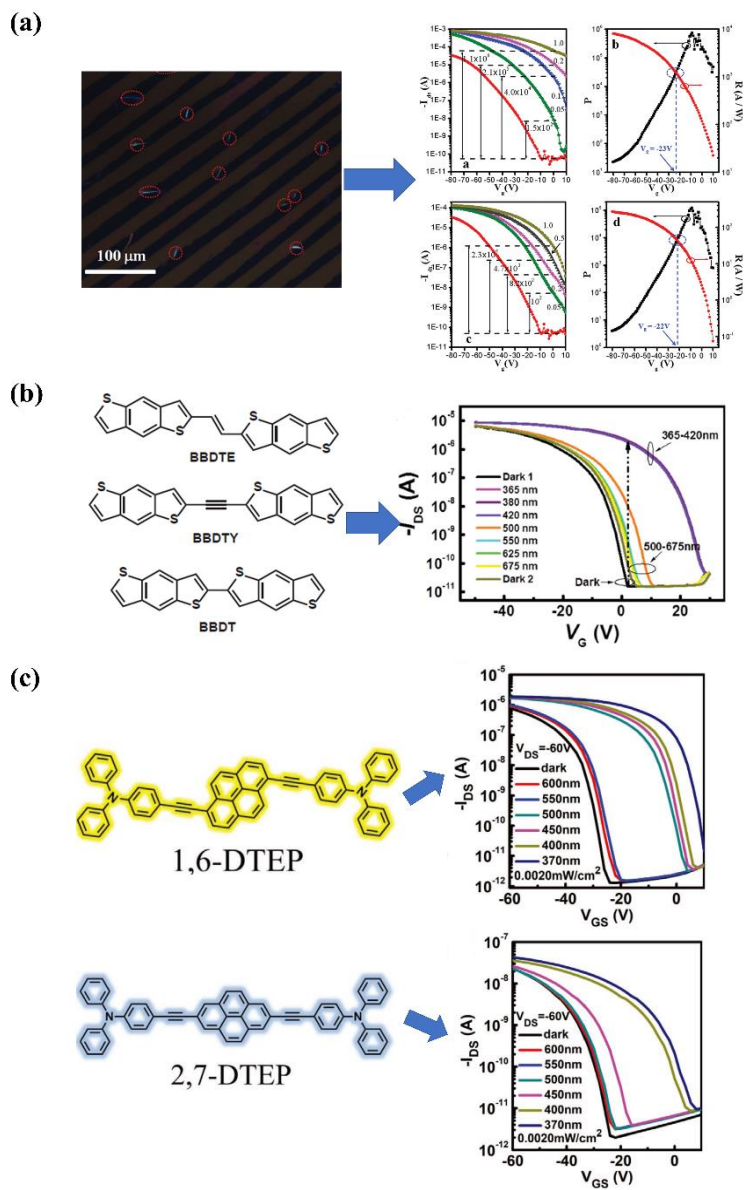

Fig. 2 Characteristics of single component OSCPTs. (a) Optical microscopy (OM) image of C8-BTBT OSCPT device and its performance. (b) Molecular structure of BBDTE and transfer curves of corresponding OSCPTs devices illuminated by light of different wavelengths. (c) Molecular structure of 1,6-DTEP and 2,7-DTEP, and transfer curves of their OSCPT devices illuminated by light of different wavelengths. Reproduced with permission from Ref. [33] (a), Ref. [34] (b) and Ref. [35] (c).

To further explore the advantages of such devices, Li and co-authors[37] fabricated OSCPTs based on BOPAnt grown by physical vapor transport (PVT) method, and thin-film transistors were also developed as a reference. The thin-film phototransistor showed a R of 9.75 A $W^{-1}$ under 1 mW $cm^{-2}$ blue LED illumination whereas the OSCPT device displayed a R of 414 A $W^{-1}$ under same conditions [Fig. 3(a)]. Furthermore, using low power monochromatic UV light with a wavelength of 350 nm, the R of 3100 A $W^{-1}$ under 0.11 mW $cm^{-2}$ and an EQE of $9.5\times10^5$ % were obtained for the BOPAnt-based OSCPTs. These results confirmed that the optical detectivity of single crystal phototransistor is much higher than that of thin-film equivalents, which is because of much longer exciton diffusion length and better carrier mobility of single crystals compared to that of thin-film device, as also evident by the larger $V_{on}$ shift towards the positive voltage direction. To further improve the optical detection performance, some researches optimized the device structure. For instance, Liu et al.[38] demonstrated an organic single crystal vertical field-effect transistors based on 2,6-diphenyl anthracene (DPA) [Fig. 3(b)]. Their fabricated devices exhibited a record-breaking high Ion/off ratio of $10^6$ with a high current density of 100 mA $cm^{-2}$ under a low voltage of −5 V. Furthermore, superior photodetection performance with an R of 110 A $W^{-1}$ and D* of $10^{13}$ Jones was also obtained under 420 nm light illumination. This type of device configuration integrates high mobility of single crystal and high output current of vertical structure, thus realizing excellent performance and improved optical detection. Li and colleagues[39] fabricated rubrene single crystal phototransistor memory devices in which the active layer was comprised of rubrene single crystal only without any additional memory functional layer. Even at a high temperature of 100 °C, the device showed a stable memory effect, indicating that deep traps act as electron acceptors. In general, the memory effect is caused by the existence of a floating gate in the dielectric layer, which can store carriers at high bias voltage, and maintain the state until erasing under reverse high bias voltage that transforms the device into another state. However, since no carrier traps layer was adopted in these devices, the authors supposed the existence of carrier traps in rubrene single crystal, which was confirmed by x-ray photoelectron spectroscopy measurements that revealed the presence of oxidation states in rubrene single crystal, which can act as traps for electrons. In conclusion, the photo-memory effect demonstrated in their work is attributed to the oxidation states in rubrene single crystal as well as to the hydroxyl groups at the semiconductor/$SiO_2$ interface.

Various methods for the preparation of organic single crystals and corresponding device parameters are summarized in Table 1. Although single component OSCPTs have many advantages, their photo-detectivity is limited by fluorescence quantum efficiency of materials and low exciton separation capability, which can be addressed by tuning the organic molecular design and employing multilayer structure.

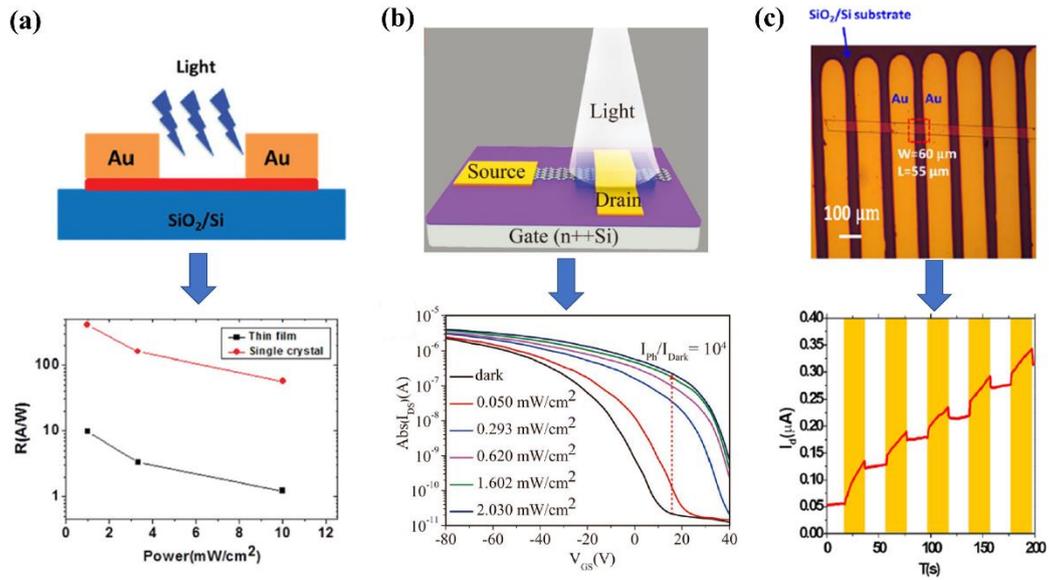

Fig. 3 (a) Device structure of a BOPAnt-based OSCPT and corresponding responsivity of thin-film transistor and OSCPTs device as a function of incident light power. (b) Vertical structure OSCPTs based on DAP and transfer curves under different light intensities. (c) OM images of rubrene OSCPTs device and its light memory switching behavior. Reproduced with permission from Ref. [36] (a), Ref. [37] (b) and Ref. [38] (c).

Table 1 Various organic single crystals, their preparation methods and corresponding device parameters

| Material | Fabrication Method | λ (nm) | Rmax (A/W) | Pmax | D* (Jones) | Ref. |
|---|---|---|---|---|---|---|
| C8-BTBT | Solution process | 365 | 1200 | $3.0 \times 10^4$ | - | [33] |
| BBDTE | PVT | 365 | 9821 | $10^5$ | - | [34] |
| 1,6-DTEP | Solution process | 370 | $2.86 \times 10^6$ | - | $1.49 \times 10^{18}$ | [35] |
| 2,7-DTEP | Solution process | 370 | $1.04 \times 10^5$ | - | $5.28 \times 10^{16}$ | [35] |
| BOPAnt | PVT | 350 | 3100 | $2 \times 10^5$ | | [36] |
| DPA | PVT | 420 | 110 | $10^6$ | $10^{13}$ | [37] |
| Rubrene | PVT | 460 | - | - | - | [38] |

### 3.2 Multilayer structured OSCPTs

Although single component OSCPTs offer long exciton diffusion length and high charge carrier mobility, the optical detection efficiency is another essential factor that depends on the exciton separation efficiency. To further improve the optical detection performance, heterojunction (HJs) structures are developed to boost the exciton separation efficiency and broaden the spectral response. Pinto and co-authors[40] reported highly efficient visible-light detection from multilayer structured OSCPTs based on rubrene single crystal and PCBM thin-film HJ [Fig. 4(a)]. The multilayer structured OSCPTs showed high hole mobility of 4-5 $cm^2V^{-1}s^{-1}$ and the device demonstrated excellent light detection capability in a broad spectral-range, from 400 to 750 nm. This HJ-based OSCPT exhibited P as high as $4\times10^4$ and R on the order of 20 A $W^{-1}$ under a light power of 27μW $cm^{-2}$, and an EQE of 52 000%; this encouraging performance is attributed to the use of HJ. In addition to using organic semiconductors to modify single crystals via developing HJs for improved performance, organic single crystals are also being used to enhance the performance of the inorganic phototransistor. For example, Russo's group[41] used rubrene single crystals to raise the optical detection capacity of graphene phototransistors. Though spectrally selective to visible wavelengths, R as large as $10^7$ A $W^{-1}$ and D* of $9\times10^{11}$ Jones was attained at room temperature for this organic-inorganic HJ based device [Fig. 4(b)], surpassing the previously reported graphene phototransistors by one order of magnitude. This improved performance is attributed to the transfer of photo-excited holes from rubrene into graphene, while electrons remain confined in the rubrene crystal that could have either originated from the built-in field at rubrene interface or generated because of the increased probability of bimolecular recombination in rubrene at high photo-excited change-carrier densities. Furthermore, Chen and co-authors[42] also used an epitaxially grown organic single crystal HJ composed of PTCDA and pentacene as the light-absorbing layer to promote photo-detectivity of graphene-based phototransistors [Fig. 4(c)], whose R can reach spectacular $10^5$ A W$^{-1}$ under low light intensity, while the graphene/PTCDA device has a maximum R~$6.3 \times 10^3$ A W$^{-1}$ over a broadband of 400−700 nm. The performance improvement is due to the built-in field in the HJ which effectively separates the generated electron−hole pairs. Liu et al.[43] further used a tandem structure based on two organic semiconductors, PTCDA single crystal and C8-BTBT polycrystalline as the light-absorbing layer to boost the performance of graphene phototransistors [Fig. 4(d)]. The performance of resulting device was improved to an R value of up to $5.76 \times 10^5$ A W$^{-1}$ with a photoconductive gain of about $1.38 \times 10^9$. Moreover, complementary light absorption spectrum of these two organic materials fulfills effective photo-detection at ultraviolet and visible range. Xu and co-workers[44] employed C8-BTBT single crystal modified by $CH_3NH_3PbI_3$ perovskite nanoparticles as the photo-active layer to fabricate phototransistors that achieved broad absorption covering the entire UV−vis range [Fig. 4(e)]. This hybrid phototransistor exhibited an ultrahigh responsivity of >$1.72 \times 10^4$ A W$^{-1}$ in the 252−780 nm region. $CH_3NH_3PbI_3$ nanoparticles in the hybrid broaden the spectral response of the pristine C8-BTBT single crystals from single-band (UV) to dual-band (UV−vis) via effective charge transfer, giving rise to a pronounced photo-response of the resulting device in the visible range. All aforementioned multilayer structured OSCPTs and their related parameters are summarized in Table 2. To sum up, developing multilayer structured OSCPTs is an efficient approach to get high photo-detectivity due to the improved exciton separation efficiency endowed by the rationally designed HJ. However, controlled preparation of single crystal HJ remains a challenging task.

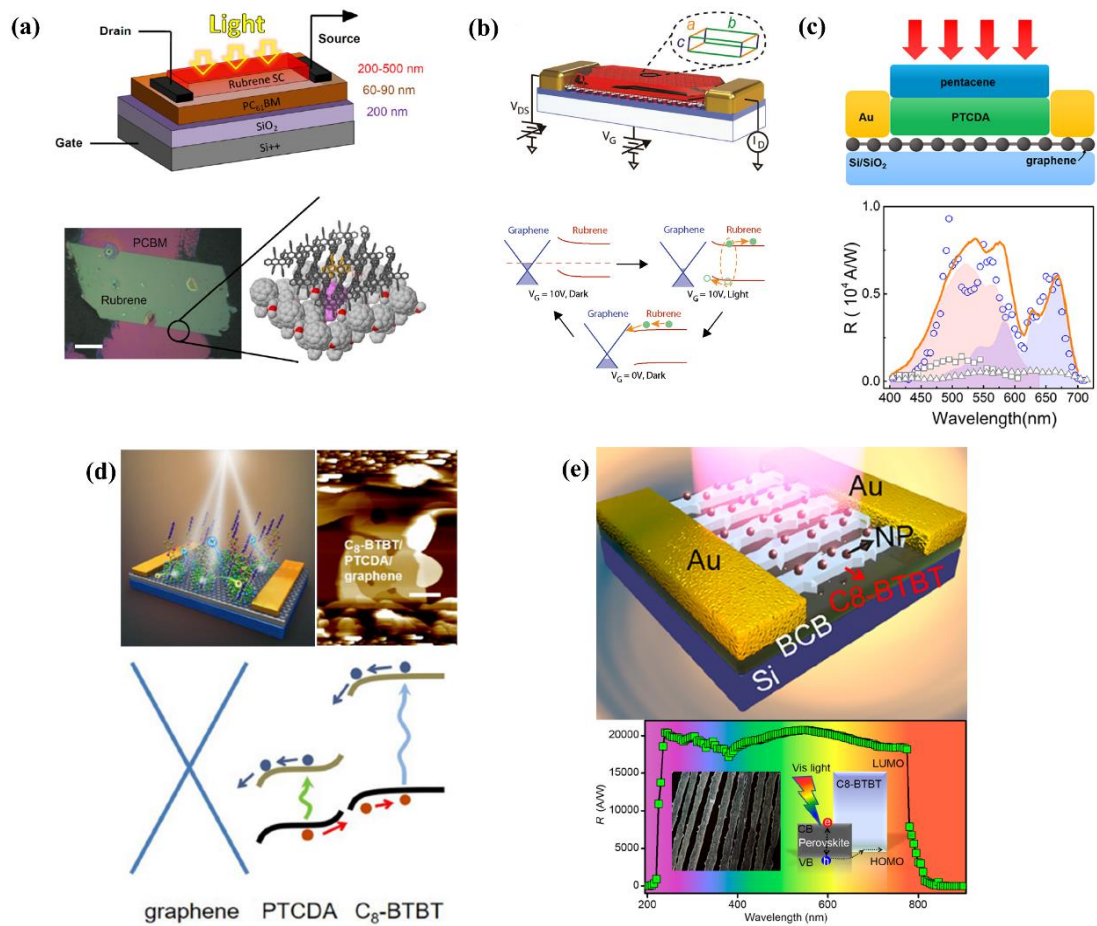

Fig. 4 (a) Device architecture and OM image of the $PC_{61}BM$/rubrene OSCPTs (Top), and molecular view of corresponding organic interface (Bottom). (b) Schematic representation of a rubrene/graphene phototransistor, and band diagrams illustrating the charge-transfer dynamics at each stage of the light modulation cycle. (c) Schematics of graphene/PTCDA crystal/pentacene phototransistors and R versus wavelength of a representative device. (d) Schematic diagram and energy band alignment of C8-BTBT/PTCDA/graphene phototransistors. (e) Schematic illustration of $CH_3NH_3PbI_3$ NPs/C8-BTBT single-crystal array-based hybrid phototransistor and R versus wavelength behavior. Reproduced with permission from Ref. [39] (a), Ref. [40] (b), Ref. [41] (c), Ref. [42] and Ref. [43].

Table 2 Key parameters of multilayered OSCPTs

| Structure | $\lambda$ (nm) | Rmax (A/W) | Pmax | D* (Jones) | Ref. |
|---|---|---|---|---|---|
| Rubrene SC/PCBM TF | 400-750 | 20 | $4 \times 10^4$ | - | [39] |
| Graphene/Rubrene SC | 500 | 107 | - | $9 \times 10^{11}$ | [40] |
| Graphene/PTCDA SC/Pentacene | 400-700 | $2.86 \times 10^6$ | $6.3 \times 10^3$ | - | [41] |
| Graphene/PTCDA SC/C8-BTBT | 355 | $5.76 \times 10^5$ | - | $1.38 \times 10^9$ | [42] |
| C8-BTBT SC/perovskite NPs | 252-780 | $1.72 \times 10^4$ | - | - | [43] |

### 3.3 Nano/Macro structured OSCPTs

In general, large-sized single crystals of most organic materials are difficult to fabricate, nevertheless, their nano/macro structured counterparts are usually much easier to prepare. Therefore, with the progress of nano/macro device processing technology, phototransistors based on nano/macro structured single crystals could now be easily fabricated. Hu's group,[45] for the first time in 2007, fabricated nano/macro structured OSCPTs based on a kind of n-type organic semiconductor, F16CuPc [Fig. 5(a)]. The resulting device displayed high performance with a maximum on/off ratio of $4.5 \times 10^4$ (at $V_G$ = –6.0 V), which could probably be due to the gate applied bias providing an efficient way for the dissociation of photo-generated excitons which is beneficial for the formation of a conducting channel in phototransistors. This work put forward a feasible strategy to realize high light sensitivity and large on/off ratio of the phototransistors and potential applications in practice. Also, Guo and co-authors[46] synthesized a new material, Me-ABT and grew micro-ribbons single crystal by solution-phase self-assembly process, then fabricated OSCPTs which showed a high mobility of 1.66 $cm^2V^{-1}s^{-1}$, a large R of 12000 A $W^{-1}$, and a P of $6 \times 10^3$ even under low light power conditions (30 mW $cm^{-2}$) [Fig. 5(b)]. Kim and co-workers[47] designed and synthesized two new anthracene-based organic molecules and obtained thin films as well as crystalline microplates using solution process [Fig. 5(c)]. OSCPTs based on individual crystalline microplates with J-aggregated molecular arrangement showed a high mobility of 0.2–1.6 $cm^2V^{-1}s^{-1}$ and an on/off current ratio of $>10^3$–$10^5$ under light irradiation. Moreover, the device showed unprecedentedly high R ($>1 \times 10^4$ A $W^{-1}$) under very low light intensity (1.4 $\mu W$ $cm^{-2}$). The high performance of their OSCPTs is mainly due to the highly ordered J-type molecular packing in crystalline microplates. Kim et al.[48] also synthesized a new pyrene-cored molecule, PY-4(THB), and fabricated single-crystalline micro-ribbon-based OSCPTs that exhibited a maximum mobility of 0.7 cm$^2$ V$^{-1}$s$^{-1}$ and an $I_{on}/I_{off}$ ratio of $4\times10^6$ [Fig. 5(d)], with R value of about 2000 A W$^{-1}$ under low light power illumination (2.6 μw cm$^{-2}$).

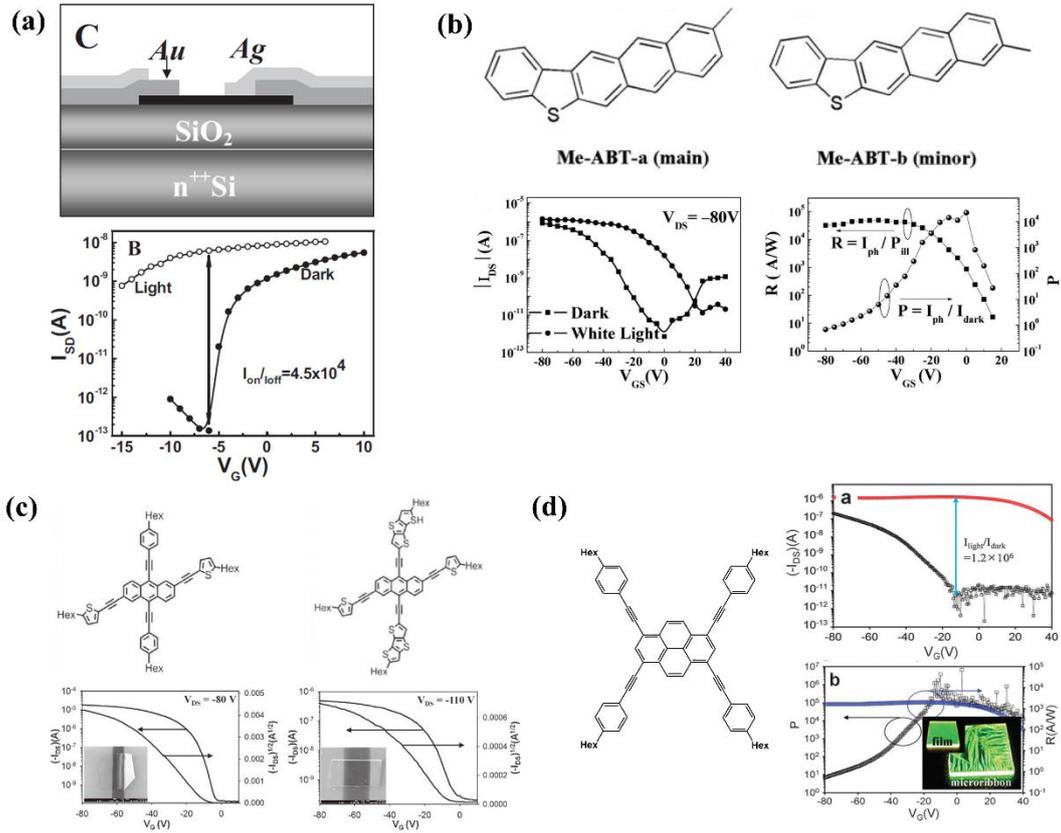

Fig. 5 (a) Schematic representation and transfer characteristics under light illumination of the F16CuPc OSCPTs. (b) Molecular structure of Me-ABT, transfer characteristics of OSCPTs based on an individual Me-ABT micro-ribbon in dark and under white light irradiation along with R and P versus $V_{GS}$ behavior. (c) Molecular structure and transfer curves of OSCPTs made of two molecules. Insets: SEM images of corresponding devices. (d) Molecular structure of PY-4(THB), transfer curves of OSCPTs under monochromatic red-light irradiation, and R and P versus $V_G$ curves. Reproduced with permission from Ref. [44] (a), Ref. [45] (b), Ref. [46] (c) and Ref. [47].

Moreover, phototransistors based on n-type organic semiconductor macro/nano structure single crystal have also been extensively studied. Yu et al.[49] fabricated

OSCPTs containing an n-type semiconductor, NID derivative single-crystalline nanowires, as the active layer (NWs) [Fig. 6(a)] and found that the single-crystalline NWs-OPTs exhibit faster charge accumulation/release rates than thin-film based counterpart. A mobility enhancement is observed when the incident optical power density increases and the wavelength of the light source matches the light absorption range of the photoactive material. Song and co-authors[50] investigated the performance of OSCPTs based on wide-band gap naphthalene diimide (NDI) derivatives, NDI-PM and NDI-TM. Their OSCPT devices exhibited high average electron mobility of 1.7 $cm^2V^{-1}s^{-1}$ and highly sensitive UV detection capability [Fig. 6(b)]. Particularly, NDI-PM based OSCPTs displayed better performance (R=7230 A $W^{-1}$, P=2.0×$10^5$, D*=1.4×$10^{15}$ Jones) than that of NDI-TM based devices (R=920 A $W^{-1}$, P=9.7×$10^2$, D*=8.3×$10^{12}$ Jones). Moreover, thin-film phototransistors were also fabricated as a reference which showed much lower electron mobility and inferior photo–detectivity, further confirming that single crystals exhibit better performance in light detection. Mukherjee et al.[51] fabricated OSCPTs based on PDI derivative microstructure single crystal, and obtained an R value of 1 A $W^{-1}$ with a P of 2.5×$10^3$ at $V_G$=12 V. Also, a high maximum R of 7 $AW^{-1}$ was achieved at a high gate bias regime ($V_G$=50 V) with an optical power of 7.5 $mWcm^{-2}$ [Fig. 6(c)]. Near-infrared sensitive phototransistors based on n-type 2D nanostructured organic single crystals were also realized by Wang and colleagues[52]. They first used solution epitaxy method to prepare 2D single crystal of TFT-CN with ultralow thickness of below 5 nm, that showed high electron mobility of 1.36 $cm^2V^{-1}s^{-1}$ and an ultrahigh on/off ratio exceeding $10^8$. Furthermore, high responsivity was achieved with a maximal R and EQE of as high as 9×$10^4$ A $W^{-1}$ and 4 × $10^6$%, respectively, which is attributed to the high quality of single crystals [Fig. 6(d)].

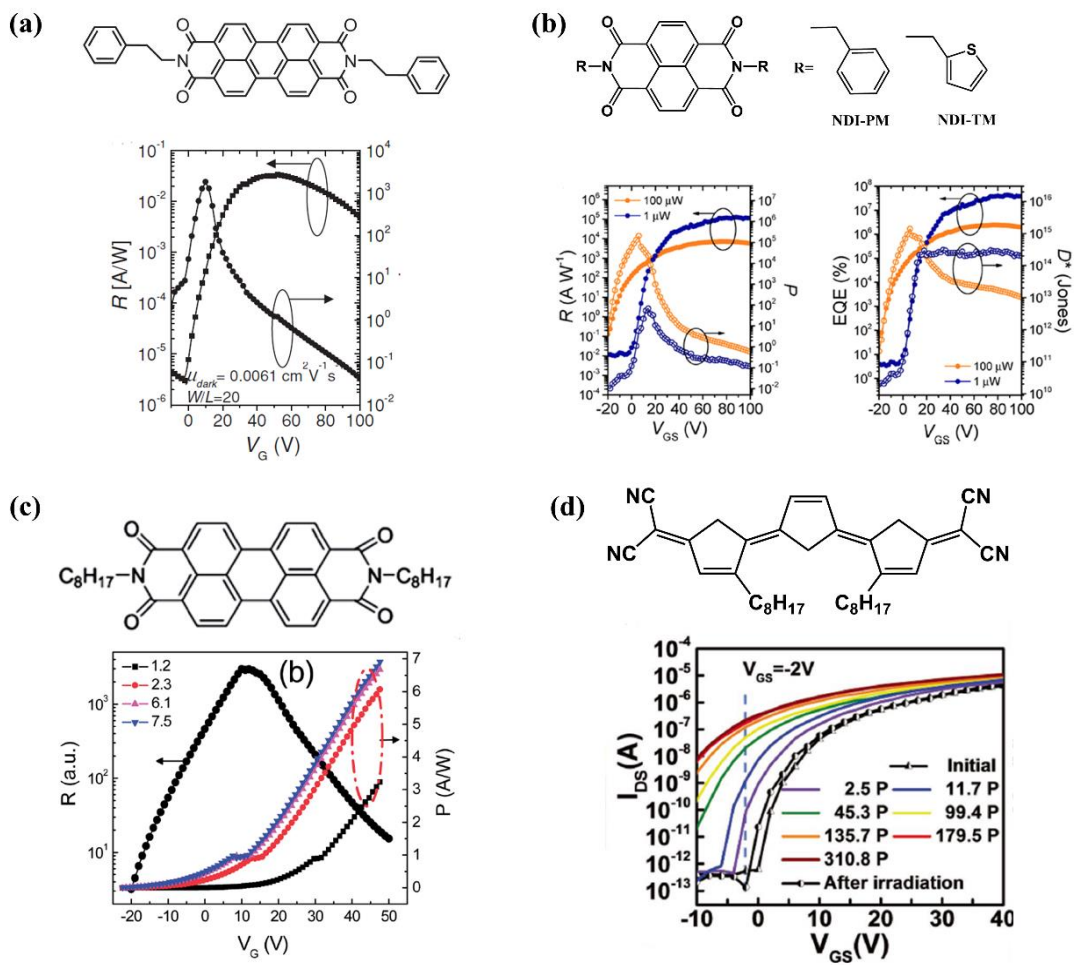

Fig. 6 (a) Chemical structure of BPE-PTCDI and R versus $V_G$ of corresponding OSCPTs device. (b) Chemical structures of NDI-PM and NDI-TM, and R, P, EQE and D* versus $V_G$ of corresponding OPSCTs devices. (c) Chemical structure of PTCDI-C8, and R and P versus $V_G$ of resulting OSCPTs device. (d) Chemical structure of TFT-CN 2D single crystal and transfer curves of resulting OSCPTs based on micrometer-sized single crystal under different power light illumination. Reproduced with permission from Ref. [48] (a), Ref. [49] (b), Ref. [50] (c) and Ref. [51].

Jiang et al.[53] used typical p-type and n-type organic single crystals, TTF and TCNQ to fabricate micro-structured single crystal phototransistor. Upon switching the light on and off, TCNQ rapidly responded within 500 ms, while the response period of TTF exceeded 60 s. Notably, TTF showed large persistent conductivity, while TCNQ did not; this phenomenon may be related with the band gaps of these materials and compactness of their molecular packing. TCNQ achieved an on/off ratio of 160, which indicates good switching effect. Furthermore, Mukherjee et al.[54] also developed a solution method to prepare periodic arrays of large, elongated crystals of TCNQ and fabricated phototransistors with polymer as the gate dielectric, and got high performance with $I_{light}/I_{dark}$ =31 at $V_G$=0.3V and R greater than 1 mA W$^{-1}$ at low driving voltages and low optical powers. Notably, same materials would grow different phase under specific conditions with diverse properties. Wang et al.[55] achieved α and β-phase micro/nanometer-sized single crystals of an organic semiconductor, BPEA via controlling the synthesis method and confirmed by single crystal diffraction results. The device with β-phase single crystals exhibited very high photo-switching performance (on/off current ratio of about $6\times10^3$), while the counterpart with α-phase displayed high field-effect performance. Obviously, the device performance of individual α and β-phase single crystals showed strong phase dependence. Single crystal nanoparticles have also been used to prepare phototransistors. Nguyen and co-authors[56] fabricated PTCDA single crystal nanoparticles with particle diameter of about 80-85 nm; the resulting phototransistors showed an electron mobility of as high as 0.08 cm$^2$V$^{-1}$s$^{-1}$ at 300K and 0.5 cm$^2$V$^{-1}$s$^{-1}$ at 80K, nearly approaching the intrinsic electron mobility of PTCDA. Further, the device also showed a fast response time with a high EQE value of $3.5\times10^6$. Moreover, other nanostructured organic semiconductors have also been used to prepare OSCPTs. Liu et al.[57] used NaT2 nanofibers single crystal to prepare OSCPTs, and attained an R of 0.91 A W$^{-1}$. In comparison, the R value of NaT2 thin-film was only $6\times10^{-3}$ A W$^{-1}$. Yao and co-authors[58] used direct photolithography method to optimize the precision of device preparation and realized a spatial resolution below 300 nm, thereby allowing the preparation of nano/macro structured single crystal devices. Two organic semiconductors, Dph-BTBT nanoflakes and PTCDI-C8 NWs were used as the active layers in phototransistors; the resulting devices showed photo-responsivity as high of $4.8\times10^4$ and photosensitivity of $2.9\times10^4$ A W$^{-1}$, respectively. Moreover, they fabricated flexible CMOS inverts based on nano/macro structured single crystal by direct photolithography. Considering the high quality of molecular crystals attained via direct photolithography, this approach would be robust toward the fabrication of large-area and flexible organic nano-electronics, and facilitate the application of sophisticated lithography for plastic electronics. The device performance parameters of nano/micro structured OSCPTs are summarized in Table 3. In short, researchers have employed nano/macro device processing technology to fabricate OSCPTs devices with enhanced performance. Although this technology has shown exciting prospects, there are still several technical issues that should be addressed for further development of this field towards commercialization, such as subsequent growth and anisotropy of organic single crystal.

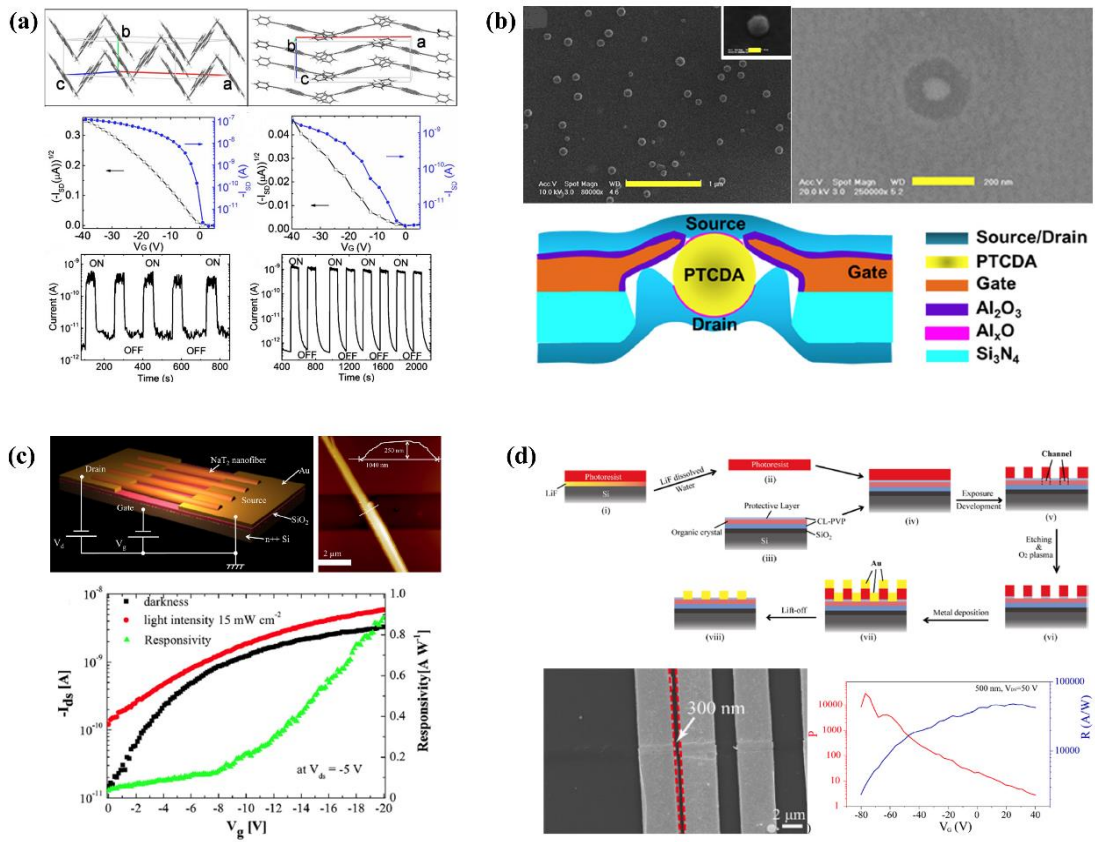

Fig. 7 (a) Single crystal diffraction results, transfer curves and photo-switching characteristics of OSCPTs based on α (right) and β-phase (left) of BPEA. (b) SEM image of PTCDA nanoparticles and schematic diagram of corresponding OSCPT device. (c) Schematic illustration of OSCPT device based on NaT2 nanofibers, corresponding AFM image and R versus $V_G$. (d) Schematic diagram of direct photolithography technique, SEM image of patterned electrodes and R versus $V_G$ of corresponding OSCPTs device. Reproduced with permission from Ref. [54] (a), Ref. [55] (b), Ref. [56] (c) and Ref. [57].

Table 3 Key parameters of nano/micro structured OSCPTs.

| Materials | Microstructure | λ (nm) | Rmax (A/W) | Pmax | D* (Jones) | Ref. |
|---|---|---|---|---|---|---|
| F16CuPc | Ribbons | 350 | | $4.5 \times 10^4$ | | [44] |
| Me-ABT | Ribbons | | 12000 | $10^4$ | | [45] |
| Anthracene derivatives | Microplates | | $> 1 \times 10^4$ | $10^3$–$10^5$ | | [46] |
| Pyrene derivatives | Microribbon | 365 | $10^4$ | $4 \times 10^6$ | | [47] |
| NID derivatives | Nanowire | Red and green | $1.4 \times 10^3$ | $4.96 \times 10^3$ | | [48] |
| NID-PM | | 365 | 7230 | $2 \times 10^5$ | $1.4 \times 10^{15}$ | [49] |
| NDI-TM | | 365 | 920 | $9.7 \times 10^2$ | $8.3 \times 10^{12}$ | |
| PTCDI-C8 | Microstructures | White | 7 | $2.5 \times 10^3$ | | [50] |
| TFT-CN | 2D structure | NIR | $9 \times 10^4$ | $10^{18}$ | $6 \times 10^{14}$ | [51] |
| TCNQ | Microrods | White | >0.001 | 31 | | [53] |
| BPEA | α-phase | White | - | $10^2$ | | [54] |
| | β-phase | | - | $6 \times 10^3$ | | |
| PTCDA | Nanoparticle | UV | - | - | - | [55] |
| NaT2 | Nanofibers | | 0.91 | | | [56] |
| Dph-BTBT | Nanoflakes | 500 | $4.8 \times 10^4$ | $2.9 \times 10^4$ | | [57] |

### 3.4 Co-crystal phototransistors

In addition to using single crystals to improve the mobility and exciton diffusion distance for enhanced detection efficiency, organic semiconductors with donor-acceptor (D-A) structure are also used to improve the exciton separation efficiency, which can lead to high detection performance. A co-crystal is a multi-component molecular crystal formed by otherwise separately stable crystalline or amorphous solids.[59] In order to combine the advantage of D-A structure and single crystals to obtain high performance, Zhang et al.[60] reported, for the first time, the preparation of D−A co-crystals based on a substituted tetrathia annulene as the donor and a C60 or C70 as the acceptor with a 2D segregated alternating layer structure by a simple drop-casting method, which displayed ambipolar transport properties as well as high photo-responsivity, as shown in Fig.8. Moreover, such a D−A blend system with a well-defined interdigitated structure could serve as an ideal model system for the investigation of charge separation and transport, which will help to thoroughly understand the physical processes involved in OPTs.

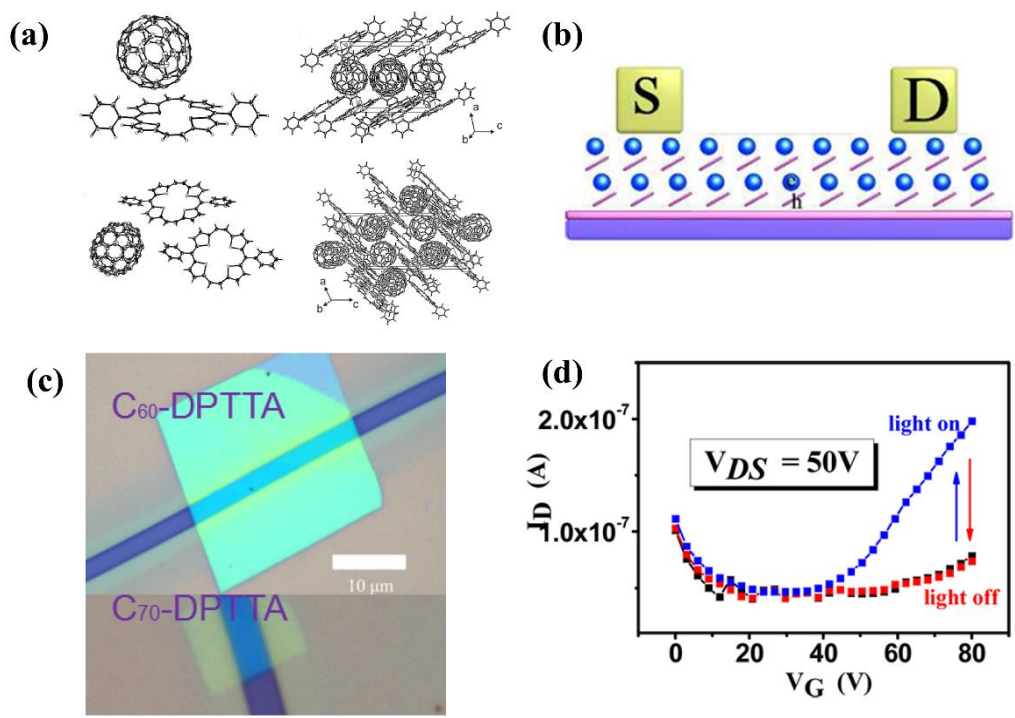

**Fig. 8** (a) Molecular structures and stacking patterns of C60-DPTTA and C70-DPTTA co-crystal. (b) Schematic diagram of C60-DPTTA or C70-DPTTA co-crystal based ambipolar transistors. (c) OM image of the device obtained with an individual C60-DPTTA or C70-DPTTA co-crystal. (d) Photocurrent response of C70−DPTTA co-crystal transistor under light irradiation. Reproduced with permission from Ref. [58]

**4. Summary and outlook**

OPTs are critical components for next-generation flexible electronics. After extensive research efforts in exploring synthesis protocols for new organic semiconductors and development of various device configurations, the performance of OPTs have been greatly improved. Among various types of OPTs, OSCPTs are promising candidates for realizing high-performance devices due to their long-range order, few defects and high mobility. In this review, we have explained the device architectures, working principles and figure of merits of OPTs, and summarized the recent approaches being adopted for the development of OSCPTs based on their classification by device structures. Despite the great achievements made in the development of OSCPTs, there remain various challenges that hinder the further improvement in the performance of these devices, like instability, slow response and difficulty in the preparation of large-area single crystals. Researchers have made great efforts to address such issues for the progress of OSCPTs towards industrialization. In the following, we provide our viewpoint to further accelerate the development of OSCPTs: (i) new organic materials with high charge carrier mobility as well as high fluorescence quantum yield need to be further developed, however, very few organic semiconductors can integrate these properties simultaneously. A potential method is to utilize high fluorescence quantum yield groups (such as triphenylamine, carbazole, acridine and so on) to modify the high mobility cores (for instance anthracene, pentacene, pyrene, Benzo[b]benzo[4,5]thieno[2,3-d]thiophene and so on), and choose appropriate linking groups to regulate the molecular aggregation in single crystals to achieve high mobility and high photo-detectivity performance in a single material; (ii) at present, the common methods of preparing single crystals include PVT and solution process, which are quite difficult to grow large and well-oriented single crystals and hence, limit the practical applications of OSCPTs. Therefore, well-controlled methods should be explored to facilely attain large-area single crystals with a particular focus on understanding the crystal growth mechanism; (iii) exciton separation efficiency is a key factor that affects the photo-detectivity, benefiting from the experience of organic photovoltaics, designing D-A HJ structure could be a potential approach to achieve high photo-detectivity and broadband light detection; (iv) co-crystal is an interesting system which is rarely used in phototransistors on account of its strict requirements of materials choice and preparation techniques, however, co-crystal formation of single crystal bulk HJ is beneficial for achieving high-performance phototransistors. Therefore, it requires special consideration to garner the full potential of this potential method in the future. To sum up, challenges and opportunities coexist in the field of OSCPTs and we believe that through continued research efforts, these crucial electronic components would be further developed towards practical applications.

**Acknowledgements**

This study is supported financially by the Key-Area Research and Development Program of Guangdong Province (2019B010924003), the Shenzhen Peacock Plan (KQTD2014062714543296), a Shenzhen Science and Technology Research Grant (JCYJ20180302153514509), the Guangdong International Science Collaboration Base (2019A050505003), the Shenzhen Engineering Research Center (Shenzhen Development and Reform Commission [2018]1410), and the Shenzhen Key Laboratory of Organic Optoelectromagnetic Functional Materials (ZDSYS20140509094114164), Natural Science Basic Research Program of Shaanxi (Program No. 2019JLP-11).